\documentstyle[12pt]{article}
\begin{document}
\begin{center}
{\large \bf MEAN FIELD DESCRIPTION OF KONDO-STABILIZED SPIN LIQUID 
IN KONDO LATTICE WITH HEAVY FERMIONS}\\
\par
{\large \bf K.A.Kikoin, M.N.Kiselev, A.S.Mishchenko}\\
{\it RRC "Kurchatov Institute", 123182 Moscow, Russia}\\             
\end{center}
\begin{abstract}
Microscopic mechanism of forming the spin liquid of neutral fermions in Kondo 
lattices with nearly integer valence of the f-electrons in presented. 
It is shown that the competition between the Kondo scattering and the indirect 
exchange interaction can result in stabilization of a state with separated
spin and charge excitations with a Fermi-type spectrum for both of them. 
The origin of strong antiferromagnetic fluctuations is briefly discussed.
\end{abstract}

\newpage
1. The well-known Doniach dichotomy \cite{1} for the ground state of the Kondo 
lattice offers the antiferromagnetic state at small values of effective 
coupling parameter $\alpha=J_{sf}/\varepsilon_F$ and the nonmagnetic 
"Kondo-singlet" state at large values of $\alpha$ as the ground states 
of the system. Here $J_{sf}$ is the on-site 
$sf$-exchange integral and $\varepsilon_F$ is the energy characterizing the 
width of conduction electron band. The critical region where two regimes 
compete is determined by the condition 
$\alpha^2_{c0}\approx\exp(-1/2\alpha_{c0})$, or, in other terms, 
$T_N\approx T_K$ where 
$T_{N,K}$ are the Neel and Kondo temperatures, respectively. We show in this 
paper that the third possibility can be realized at 
$\alpha\sim\alpha_{c0}$: 
the Kondo-stabilized spin liquid state of the RVB type is formed instead of 
the Kondo singlet state. This neutral spin-fermi liquid interacts strongly 
with charged conduction electrons, the low-temperature termodynamics is 
determined mainly by the spin-component of this two-component Fermi liquid, 
whereas the conduction electrons with enhanced effective masses are 
responsible for the charge transport and diamagnetic properties of 
the system in accordance 
with the phenomenological picture offered in \cite{2,22,222}. 

The idea of Kondo-type stabilization mechanism of the resonance valence bond 
state was offered in \cite{3}.
Coleman and Andrei described a picture of 
resonating valence bonds linked up with the electrons in 
a layer of $T_K$ around the Fermi energy. 
However, the introduction of the Kondo-type mean field resulted in 
charge transfer from the low-energy conduction electrons to the neutral 
fermi excitations originating from the f spins.
Thus they obtained a "charged 
spin liquid" like that considered in the mean-field slave-boson theory 
\cite{9}.
 The general arguments against 
such a procedure were presented in \cite{2}, and here we show that  
formation of spin liquid can prevent the system from forming 
the Kondo singlets, and the Fermi liquid with neutral spin and charged 
electron components is formed instead. 

2. We start with a standard Kondo lattice Hamiltonian $H_{eff}$ which can be 
derived from the general Anderson lattice model in the case of nearly integer 
valence.  
\begin{equation}
H_{eff}=\sum_{{\bf k} \sigma}\varepsilon_k c^+_{\bf k \sigma}c_{{\bf k}\sigma} 
+J_{sf}\sum_{\bf i} {\bf s}_{\bf i} {\bf S}_{\bf i}\; .
\label{1}
\end{equation}
Here $\varepsilon_k$ is the band level of conduction electron, 
${\bf S}_{\bf i}$ and 
${\bf s}_{\bf i}=c^+_{{\bf i}\sigma} \hat{\sigma} c_{{\bf i} \sigma}$
are  the operator of localized f spin and conduction electron spin,
respectively, $\hat{\sigma}$ is the Pauli matrix.

To consider this strongly coupled electron liquid in a controllable way, 
we start from the high temperature region $T>T_K$ where the non-crossing 
approximation (NCA) is acceptable and logarithmic perturbation theory 
is valid. We study the competition between the homogeneous RVB 
\cite{3,6} and Neel ordering within 
the mean field approximation by using the Abrikosov pseudofermion operator 
representation (see, e.g., \cite{8}) for the f-spins,
${\bf S}_{\bf i} = 
f^+_{{\bf i} \alpha} \hat{\sigma}_{\alpha \beta} f_{{\bf i} \beta}$. 
Then the on-site and intersite contributions of the indirect RKKY interaction 
to the self energy part of the pseudofermion temperature Green's function  
$F(i\epsilon_n)$ are given by the diagrams of fig.1a and 1b, 
respectively. 
Here the dashed and full lines stand for the pseudofermion and conduction 
electron propagators, hatched squares and crosshatched rectangles 
symbolize the sf- and ff vertex parts 
renormalized by the multiple on-site electron-spin scattering. 
First of these diagrams describes the standard molecular field due to the 
indirect RKKY interaction acting on the spin in a site ${\bf i}$, and second 
one corresponds to the "anomalous" RVB-type correlations. 
The polarization operator in the diagram 1a is taken in a nearest neighbor 
approximation, 
\begin{equation}
\Pi_{\bf q}(i\omega_n)=\Pi_{R}(i\omega_n)S({\bf q})
\label{1a}
\end{equation}
where $S({\bf q})=\sum_{\bf l}\exp i{\bf q}{\bf R}_{jl}$ is a structure factor, 
$R_{jl}=|R_{\bf j}-R_{\bf l}|$. The possibility of commensurate magnetic order 
with the antiferromagnetic vector ${\bf Q}$ such that ${\bf QR}=\pi$ and 
$J(R)\equiv J_{\bf ij}=J_{sf}^2\Pi_R(0)$ is suggested. 

We consider the 
renormalization of these diagrams by the on-site Kondo scattering at $T>T_K$. 
According to the NCA approach, this 
scattering processes renormalize independently the vertices 
$\Gamma$ belonging to different cites $({\bf i,j})$, and this renormalization 
should be taken in the logarithmic approximation where the maximum energy 
variable in the argument of the vertex part is retained. This energy is 
determined by the characteristic frequencies of the electron Green's function 
$G(i\varepsilon_m)$ entering the polarization operator 
$\Pi_R(i\omega_n)$
in the RKKY interaction.  

Within the mean-field approach the transition temperatures 
$T_N$ and $T_{RVB}$ are found from the self-consistent equations for 
the temperature Green's functions with the self-energy parts shown 
in figs.1({\it a,b}). 
These approximations are obtained in a static approximation for RKKY 
interaction.
It was found by Doniach \cite{1} that the Kondo scattering 
destroys the magnetic order in a Kondo lattice at least at 
$\alpha > \alpha_c$, 
and the "Kondo singlet" state can be formed in this strong coupling 
limit. Later it was recognized \cite{3} 
that one can expect the stabilization of RVB phase instead of the hypothetical 
Kondo singlet phase provided the inequality 
$|T_{RVB}-T_{N}|\ll T_K$ is valid.  
Here we shall see that this inequality is realized in a 
critical region  $\alpha^{_{_>}}_{^\sim} \alpha_c$ of the Doniach's phase 
diagram where $T_N\approx T_K$ and the Kondo scattering nearly screens out 
the localized moments thus reducing the Neel molecular field 
\begin{equation}
B_N(T)=\lambda_1\tilde{J}(R,T)\langle S_z\rangle 
\label{1b}
\end{equation}
Here $\lambda_1$ is the numerical 
factor determined by the lattice geometry.

To find the molecular field $B_N(T)$ renormalized by the Kondo scattering 
one should first estimate the effective 
exchange integral $\tilde{J}(T)$ entering the self-energy parts of 
Figs 1({\it a,b}) 
(the hatched squares),  
\begin{equation}
\tilde{J}(R,T) =T\sum_n G^2(R,\varepsilon_n)\Gamma^2(\varepsilon_n,T)
\label{10}
\end{equation}
where the asymptotic expression for the Green's function in a coordinate 
representation for $p_FR\gg 1$ and the spherical Fermi surface has the form 
\begin{equation} 
G(R,\varepsilon)= -\frac{v_F}{2\pi p_fR}\exp(-\frac{|\varepsilon|R}{v_F}
+ip_FR\;{\rm sgn}\varepsilon)\;,
\label{101}
\end{equation} 
the vertex part 
\begin{equation}
\Gamma(\varepsilon)=J_{sf}
\left(1-2\alpha\ln\frac{\varepsilon_F}{\max\{\varepsilon, T\}}\right)^{-1}
\label{1010}
\end{equation}
is the Kondo-enhanced sf-exchange integral.  The simple calculation give 
\begin{equation}
\tilde{J}(R,T)\approx \Phi(p_FR)\alpha^2\varepsilon_f
\int_T^{\varepsilon_F}d\varepsilon
\frac{ exp(-\frac{\varepsilon}{\varepsilon_F}p_FR) }
{\left[1+2\alpha\ln \frac{\varepsilon}{\varepsilon_F} \right]^2}  
\label{11} 
\end{equation}
where $\Phi(x)$ is the oscillatory RKKY function. 

In spite of the enhancement of the indirect exchange at $T>T_K$, the 
Kondo scattering eventually suppreses the AFM ordering because the same 
Kondo processes screen the localized spin entering the molecular field 
(\ref{1b}). The screening factor (the insert $K(T)$ at fig.1b)   
has the form 
\begin{equation} 
K(T)=1-2\alpha\ln\frac{\varepsilon_F}{T}/\ln\frac{T}{T_K} 
\label{102}
\end{equation}
within the 
logarithmic perturbation theory (see, e.g., \cite{7}) and formally turns 
into zero at 
$T^*=T_K(\frac{\varepsilon_f}{T_K})^{\frac{\alpha}{1+2\alpha}}>T_K$. However,
this function should be corrected for the exact solution of the on-site 
Kondo problem \cite{8}, so the local spins are only partially screened 
at $T$ approaching $T_K$,  $[K(T_K)\approx 0.37]$. Nevertheless, the molecular 
field is diminished essentially in comparison with the pure Heisenberg exchange 
$J_{\bf ij}$. 

The spin  screening is absent for the RVB molecular field $B_{RVB}$. To find 
this field one should calculate the self-energy part of Fig.1b in a static 
approximation $\Pi_{\bf q}(i\omega_n)\rightarrow \Pi_{\bf q}(0)$. Then the 
field $B_{RVB}$ is given by the equation (see, e.g., \cite{3,6,5})
\begin{equation}
B_{RVB}(T)=\lambda_2\tilde{J}(R,T)\Delta_{RVB}
\label{10101} 
\end{equation}
where 
$\Delta_{RVB}=\sum_{\sigma}<f^+_{{\bf i}\sigma}f_{{\bf j}\sigma}>$, 
is the "anomalous" RVB correlator.
The homogeneous RVB state  
possesses the fermi-type excitation spectrum with a dispersion 
$u(k)=\lambda_2\tilde{J}(R,T)\Delta_{RVB}S({\bf k})$ 
To find the mean-field transition temperature $T_{RVB}$ one should solve a 
self-consistent equation 
\begin{equation}
\Delta_{RVB}=(zN)^{-1}\sum_{\bf k}S({\bf k})
\tanh \frac{\lambda_2\tilde{J}(R,T)\Delta_{RVB}S({\bf k})}{2T} 
\label{12}
\end{equation}
(cf.\cite{3}). 
Here $z$ is the coordination number. 
If the lattice 
geometry is such that several coordination spheres $(m)$ give 
$\Phi(k_F R^{(m)})$ of negative sign and comparable values, then the RVB 
molecular field has the form \cite{5}
\begin{equation}
B_{RVB}(T)=\lambda_2 \Delta_{RVB} \sum_m \tilde{J}(R_m,T)
S^{(m)}({\bf k})\;.
\label{14}
\end{equation}
where $m$ is the number of coordination sphere. To simplify the numerical 
calculations we used the approximation 
$B_{RVB}(T)\approx \tilde{\lambda}_2\tilde{J}(R_1,T)
\Delta_{RVB}S^{(1)}({\bf k})$
where $\tilde{\lambda}_2$ is a numerical factor. 

Then, the molecular fields $B_N(T)$ and $B_{RVB}(T)$ are given by 
\begin{eqnarray}
B_N(T)\simeq B_N^{0}(T)K(T/T_K)\ln^{-n}(T/T_K),\\
\nonumber
B_{RVB}(T)\simeq B_{RVB}^{0}(T)\ln^{-n}(T/T_K)\;.
\label{13}
\end{eqnarray}
where index (0) stands for stands for a "pure" Heisenberg interaction 
$J_{\bf ij}$ without Kondo renormalization. In   
the convolution procedure of eqs.(\ref{10}),(\ref{11}) 
the logarithmic enhancement is 
partially "integrated out", and $n$ is close to 1 for the reasonable values 
of the model parameters.
When comparing these equations we find that exchange enhancement favors both 
Neel and RVB states but the spin screening influences only Neel field 
thus weakening the trend to the AFM ordering. 
The possible contribution of several coordination spheres in  
$B_{RVB}$ (\ref{14}) is an additional factor stabilizing the 
spin-liquid state, because the staggered Neel field possesses the definite 
antiferromagnetic vector ${\bf Q}$. We have taken this possibility 
into account by varying the factor $\tilde{\lambda}_2$ in $B_{RVB}$. 

The phase diagram ($T_N, T_{RVB}$ vs $\alpha$) depends also on the form of 
the electron Fermi surface which influences the coordinate dependence of the 
function $\Phi(x)$. We calculated the transition temperatures for the 
spherical and cylindrical Fermi surfaces with 
$\Phi(x)\approx \pi x^{-3}\cos 2x$ and $\Phi(x)\approx -2x^{-2}\sin 2x$, 
respectively, and found that the region of existence of the RVB phase is 
larger in the latter case which is close to the real situation in 
CeRu$_2$Si$_2$ \cite{4}. 

The example of numerical solution of the mean field equation
for $T_N$,$T_{RVB}$ with cylindrical Fermi surface is presented in figure 2. 
Here again the index "0"  marks the transition temperatures 
for the bare Heisenberg interaction,
$T_N^{0},T_{RVB}^{0} \sim \varepsilon_F \alpha^2 \Phi(p_FR)$

It is known that $T_N^{0}>T_{RVB}^{0}$ in a 3D case and the RVB state does not 
arise in three dimensions, but here we find the the Kondo-processes change  
essentially the Doniach's diagram ($T_N^0,T_K$ vs $\alpha$) \cite{1} 
in a critical region 
$\alpha \sim \alpha_c$. We have instead a modified diagram 
($T_N,T_{RVB}$ vs $\alpha$) with a wide region of spin liquid phase overcoming 
the magnetic phase and arising {\it instead} of the Kondo-singlet phase. 

The critical region is characterized by the temperature sequence 
$T_{RVB}>T_N>T_K$, all of them being of the same order. This means, firstly, 
that the spins entering the RVB pairs are screened essentially by the 
Kondo interaction, and, secondly, that the neutral spin liquid is close to 
the antiferromagnetic instability. On the other hand, the Kondo scattering is 
"quenched" at $T\approx T_{RVB}>T_K$. This means that the 
Kondo singlet state is not formed in this case, so the Kondo mean field 
$\langle c^+_{i}f_{i}\rangle$ 
which could glue together the spin and charge excitations \cite{3,9} 
equals zero. Then, 
the Kondo temperature is no more the singular point of the theory, and we 
can transit to the low-temperature region $T<T_K$ where the two-component 
Fermi liquid occurs. This liquid contains the slow electrons with the energy 
$\varepsilon<T_K$ which interact strongly with the neutral spin fermions 
with the energy scaled by $T_{RVB}$. The coupling constant can be estimated 
as $\sim \tilde{J}(T_{RVB})$. Just this picture was offered in \cite{2,22,222}
for description of the heavy fermion state in the rare-earth intermetallides 
with nearly integer valence (see also \cite{12,13}. 

3. In conclusion, we obtained that the Kondo processes in Kondo lattices at 
$T$ above $T_K$ prevent the system from forming the Neel state in a critical 
region $\alpha\sim\alpha_c$ of the Doniach diagram. The same processes 
stabilize the spin liquid state which, in turn, results in binding the 
spins in the RVB pairs and quenching the Kondo processes at $T\sim T_{RVB}$. 
At low temperature we have a two-component Fermi liquid of strongly 
interacting spin and charge fermions instead of the Kondo singlet, and  
$T_{RVB}$ plays the role of characteristic temperature of the heavy fermion 
behavior. However, since $T_{RVB}$ is close to $T_N$, 
the AFM fluctuations should
be strong in this spin liquid, and the magnetic order can be restored under
small external influences. 

There are many essential effects which are left beyond the mean-field 
approximation. First of all, the mean-field approximation introducing the 
average $\Delta_{RVB}$ violates the gauge invariance of the Hamiltonian 
(\ref{1}) (see, e.g., \cite{14}). 
The fluctuations of the phase of pseudofermions should result 
in transformation of an abrupt phase transition described by equation   
(\ref{12}) into a smooth crossover. It is known \cite{14,6}
that the homogeneous fluctuations of the gauge field in a 2D $t-J$ model 
(chirality fluctuations) result in infrared singularity of the gauge field 
fluctuation correlator. In a 3D sf-exchange model the RVB transition is close
to the antiferromagnetic instability according to the phase diagram of 
fig.2, hence the phase fluctuations should be essentially inhomogeneous. 
Thus, we believe that the influence of the gauge fluctuations 
is not so destructive for the RVB pairs as in the 2D case, and the spin-charge 
separation persists in the heavy fermion systems at low temperatures.

When the retardation effects (the frequency dependence of the polarization 
operator) are taken into account in the diagram 1b, the "polaron 
renormalization" $\sim \alpha^2\omega\ln\frac{\varepsilon_F}{T}$
and the damping $\sim \alpha^2T$ of the spin-fermion spectrum arise 
at $T<T_{RVB}$. The spin fermions, in turn, should influence noticeably 
the electron spectrum \cite{2,22,13,15}. One should hope, however, that 
these processes will not change in principle the picture of  forming the 
two-component quantum liquid consisting of charged electrons and neutral 
spin fermions with different effective masses which share the responsibility 
for the thermodynamic and transport properties of the heavy fermion systems. 

This research is supported by the International Science Foundation 
(grant MBH000) and the Netherlands Organization for Scientific Research 
(grant NWO-07-03-002).

\par
{\Large \bf Figure captions} \\ \mbox{} \\
\par
Fig.1. Self-energy parts for the pseudofermion Green's functions 
$G_{\bf ii}$ (a) and $G_{\bf ij}$ (b). \\ \mbox{} \\
\par
Fig.2. Modified Doniach diagram for competing Neel and RVB phases calculated 
with the values of $z=6$, $\lambda_1/\lambda_2=2.1$, $p_FR=2.88$. 
$\alpha_{c0}$ and $\alpha_c$ are the critical points where the AFM solution 
disappears in Doniach's and modified Doniach's pictures, respectively. 

\newpage 
\begin{thebibliography}{99} 
\bibitem{1} S.Doniach, Physica B 91 (1977) 231.
\bibitem{2} Yu.Kagan, K.A.Kikoin and N.V.Prokof'ev, Physica B 182 (1992) 201.
\bibitem{22} Yu.Kagan, K.A.Kikoin and N.V.Prokof'ev, JETP Letters 56 (1992) 219. 
\bibitem{222} Yu.Kagan and N.V.Prokof'ev, J. Phys.:Cond. Matt. 5 (1993) 6189.
\bibitem{3} P.Coleman and N.Andrei, J. Phys. Cond. Mat. 1 (1989) 4057.
\bibitem{9} A.Millis and P.Lee, Phys. Rev. B 35 (1986) 3394.
\bibitem{6} P.A.Lee and N.Nagaosa, Phys. Rev. B 46 (1992) 5621.
\bibitem{8} A.M.Tsvelik and P.B.Wiegmann, Adv. Phys. 32 (1983) 453. 
\bibitem{7} A.A.Abrikosov and A.A.Migdal, J. Low Temp. Phys. 3 (1970) 519.
\bibitem{5} T.Tanamoto, H.Kohno and H.Fukuyama, J. Phys. Soc. Jpn. 62 (1993) 617.
\bibitem{4} H.Yamagami and A.Hasegawa, J. Phys. Soc. Jpn. 62 (1993) 592. 
\bibitem{12} A.Tsvelik, Phys. Rev. Lett. 69 (1992) 2149. 
\bibitem{13} P.Coleman, E.Miranda and A.Tsvelik, Phys. Rev. Lett. 70 (1993) 2960.
\bibitem{14} B.L.Ioffe and A.I.Larkin, Phys. Rev. B 39 (1989) 2960.
\bibitem{15} G.M.Eliashberg, JETP Letters, 45 (1987) 35.
\end{thebibliography}
\end{document}